\documentclass[10pt,conference]{IEEEtran}
\IEEEoverridecommandlockouts
\usepackage{cite}
\usepackage{amsmath,amssymb,amsfonts}
\usepackage{algorithmic}
\usepackage{graphicx}
\graphicspath{{figures/}}
\usepackage[spaces,hyphens]{xurl}
\usepackage{textcomp}
\usepackage{xcolor}
\def\BibTeX{{\rm B\kern-.05em{\sc i\kern-.025em b}\kern-.08em
    T\kern-.1667em\lower.7ex\hbox{E}\kern-.125emX}}

\begin{document}

\title{An Exploratory Study of Project Activity Changepoints in Open Source Software Evolution}

\author{\IEEEauthorblockN{James Walden}
\IEEEauthorblockA{\textit{Department of Computer Science} \\
\textit{Northern Kentucky University}\\
Highland Heights, KY USA\\
waldenj@nku.edu}
\and
\IEEEauthorblockN{Noah Burgin}
 \IEEEauthorblockA{\textit{Department of EE and Computer Science} \\
    \textit{University of Tennessee}\\
Knoxville, TN USA\\
noah22@vols.utk.edu}
\and
\IEEEauthorblockN{Kuljit Kaur}
\IEEEauthorblockA{\textit{Department of Computer Science} \\
\textit{Guru Nanak Dev University}\\ 
Amritsar, India\\
kuljitchahal.cse@gndu.ac.in} 
}
\maketitle

\begin{abstract}
    To explore the prevalence of abrupt changes (changepoints) in open source project activity, we assembled a dataset of 8,919 projects from the World of Code. Projects were selected based on age, number of commits, and number of authors. Using the nonparametric PELT algorithm, we identified changepoints in project activity time series, finding that more than 90\% of projects had between one and six changepoints. Increases and decreases in project activity occurred with roughly equal frequency. While most changes are relatively small, on the order of a few authors or few dozen commits per month, there were long tails of much larger project activity changes. In future work, we plan to focus on larger changes to search for common open source lifecycle patterns as well as common responses to external events.
\end{abstract}


\section{Introduction}

We performed an exploratory study of changepoints in open source project activity during the MSR 2021 hackathon. Changepoints are abrupt changes in the mean, variance, or other statistics of time series. Such changes can be caused by internal factors, such as a project moving into a maintenance phase after release of a new version, or external factors, such as discovery of security flaw that requires significant code changes to fix. Our goal in this paper is to determine the prevalence of changepoints in open source activity and to assemble a dataset of time series with changepoints that can be used to identify patterns and causes of changepoints.

We analyzed project activity time series obtained from the World of Code~\cite{ma2019world}, an archive cross-referencing over 120 million git repositories from multiple forges. We selected 8,919 projects from the World of Code that had sufficient historical data to compute monthly time series of project activity. Activity metrics included the number of commits and number of unique authors making commits per month.

Lehman's laws of software evolution~\cite{lehman1996laws} describe how time series that describe characteristics of software, such as complexity or functionality, evolve in the long run. However, these laws do not address the question of whether such time series are smooth or punctuated by changepoints. 

A five stage model of the software lifecycle has been proposed to explain how project activity changes throughout the lifecycle of a project~\cite{rajlich2000staged}. The model was adapted to account for multiple phases of growth and stabilization found in open source software evolution~\cite{capiluppi2007adapting}. These papers visually identified project phases in the time series of a few projects and did not use statistical tests, such as changepoint detection.


In contrast, we analyzed thousands of projects using a changepoint detection algorithm~\cite{van2020evaluation} to measure the prevalence and size of changepoints in open source software evolution. The two research questions for this exploratory study were:
\begin{enumerate}
    \item How common are changepoints in open source project activity?
    \item What are the sizes and magnitudes of changes at changepoints?
\end{enumerate}

\section{Data}

In order to have sufficient data for changepoint analysis, we selected open source projects with at least 4 years of history, 50 authors, and 5000 commits. We identified 8,919 projects that met these criteria in version S of World of Code project data. Projects were found in the MongoDB \texttt{WoC.proj\_metadata.S} collection.

During the course of the multi-week virtual hackathon, World of Code (WoC) data transitioned from version R to version S. We adapted data collection scripts and procedures written for version R to use the new version, in order to gain access to the new \texttt{rootfork} field it provided. Forges like GitHub contain many forks of popular projects, making it difficult to identify the repository that is used by the project team for development. Prior to version S, the only measure of centrality in a cluster of projects was algorithmically determined within WoC. The \texttt{rootfork} field identifies the true root project based on data provided by GitHub.

We collected two monthly time series for each project based on the numbers of commits and active authors. Time series were computed using the \texttt{getValues} commands that access data in pre-computed maps and tables within WoC. To get all commits for a selected project, we used the \texttt{p2c} map. We then used the \texttt{c2ta} map to retrieve the timestamp and author of each commit. A python script grouped commits by month, counting the number of commits per month and the number of unique authors who made those commits. Running these processes on World of Code servers took four days. As other hackathon projects were simultaneously using these servers, it may be possible to compute the time series in a shorter amount of time.


\section{Changepoint analysis}

As our time series data was not normally distributed, we used the nonparametric PELT (Pruned Exact Linear Time) algorithm as implemented in version 1.0.2 of the \texttt{changepoint.np} R package~\cite{killick2014changepoint}. We used the algorithm's default parameters, with the exception of specifying the minimum segment length to be three months, as we wanted to find changes in activity that lasted longer than a single month.


We found that the vast majority (over 99\%) of projects had changepoints in both author and commit time series. Figure~1 displays the distribution of changepoints in the authors per month time series. The number of changepoints is given on the y-axis, while the number of projects whose author time series have that many changepoints is displayed on the x-axis. The median number of changepoints per project was three, with most projects (94\%) having between one and six changepoints. Only 55 projects had no changepoints in their author time series. There are also a few outliers on the high side, with seven projects having ten or more changepoints.
\begin{figure}[ht!]
    \centering
    \includegraphics[width=\linewidth]{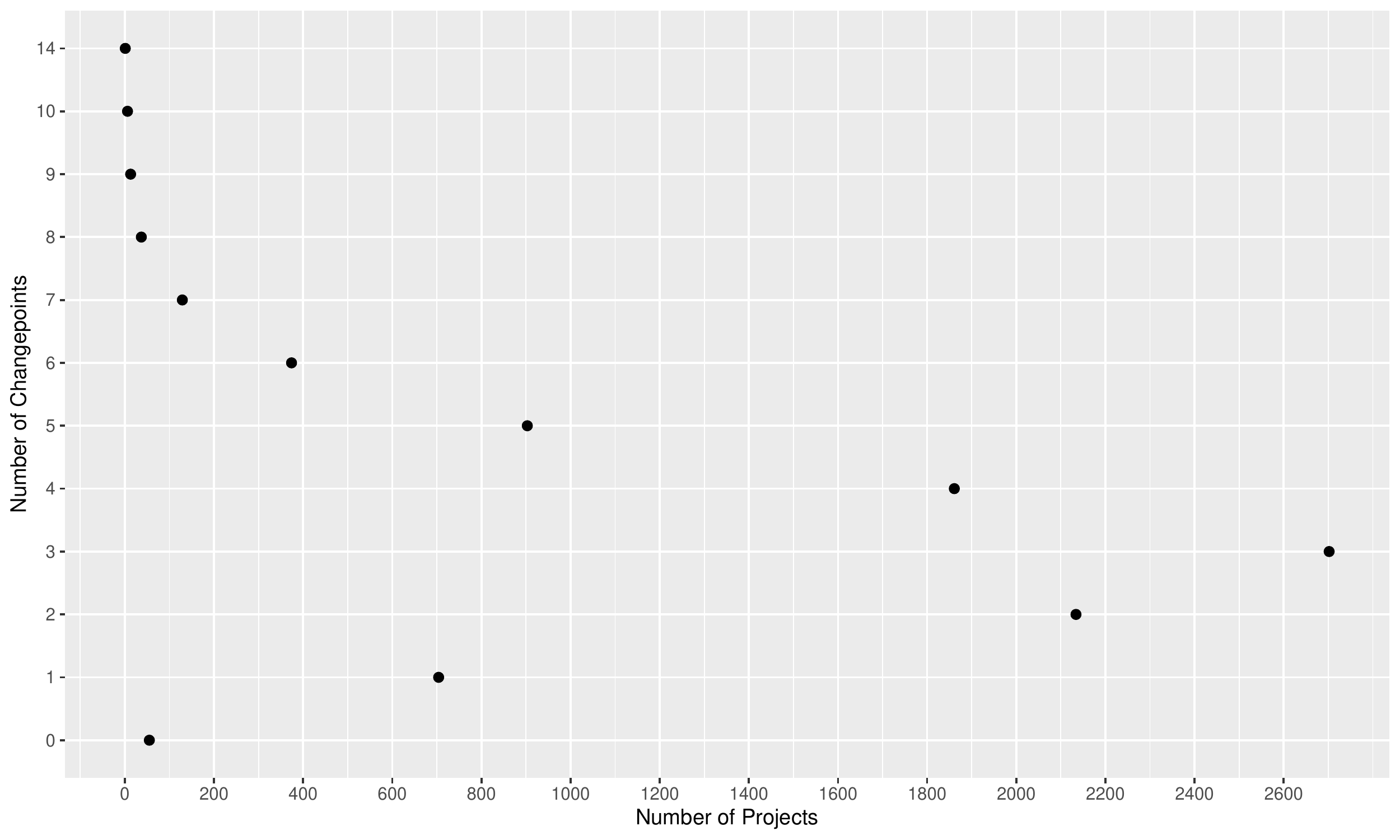}
    \caption{Number of Changepoints in Author Time Series}
    \label{fig:authorcpts}
\end{figure}

Figure~2 shows the distribution of changepoints for the commits per month time series.
Over (90\%) of projects had between one and six changepoints, and the median number of changepoints was three. There were 27 projects with ten or more changepoints, while 32 projects had no changepoints.

\begin{figure}[ht!]
    \centering
    \includegraphics[width=\linewidth]{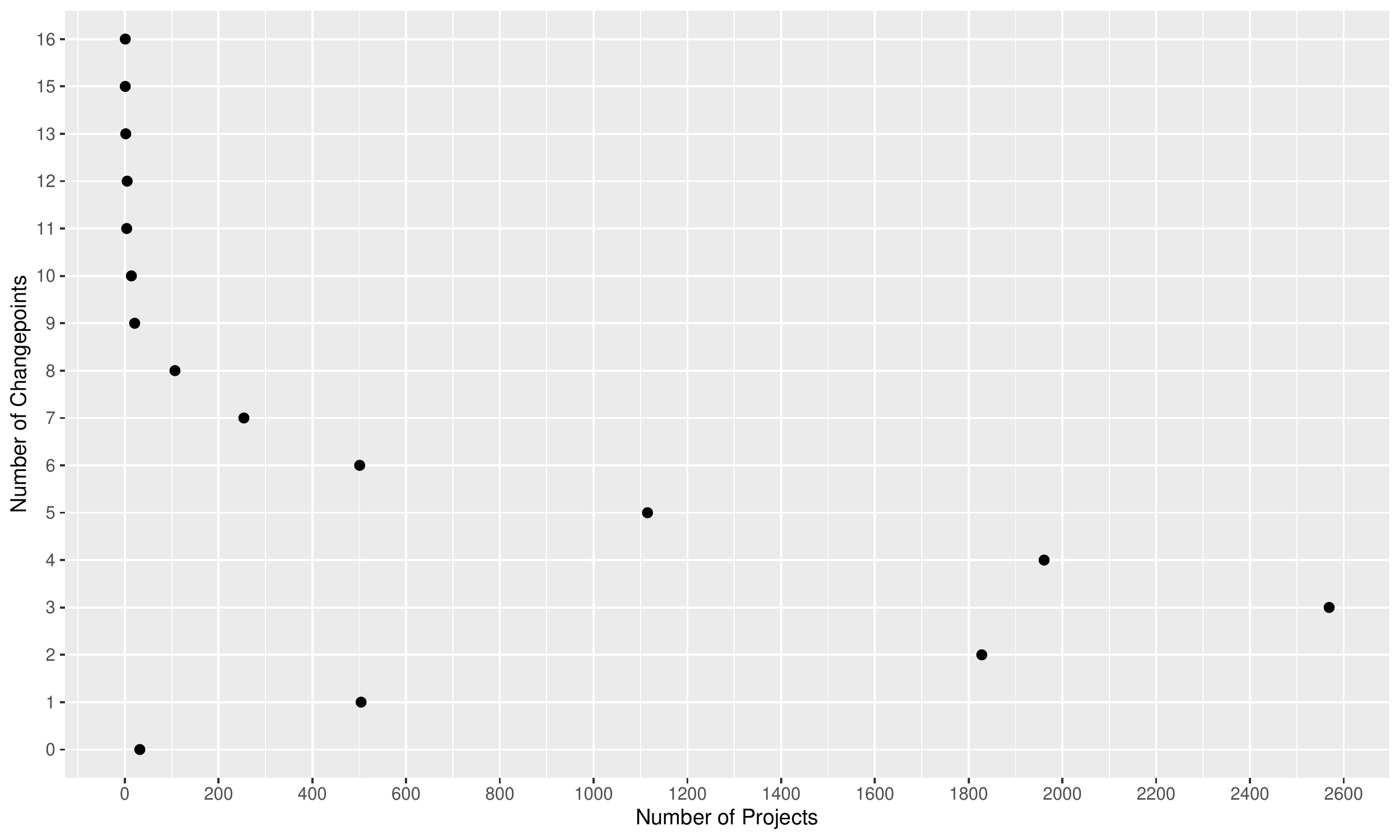}
    \caption{Number of Changepoints in Commit Time Series}
    \label{fig:commitcpts}
\end{figure}

\begin{figure}[ht!]
    \centering
    \includegraphics[width=\linewidth]{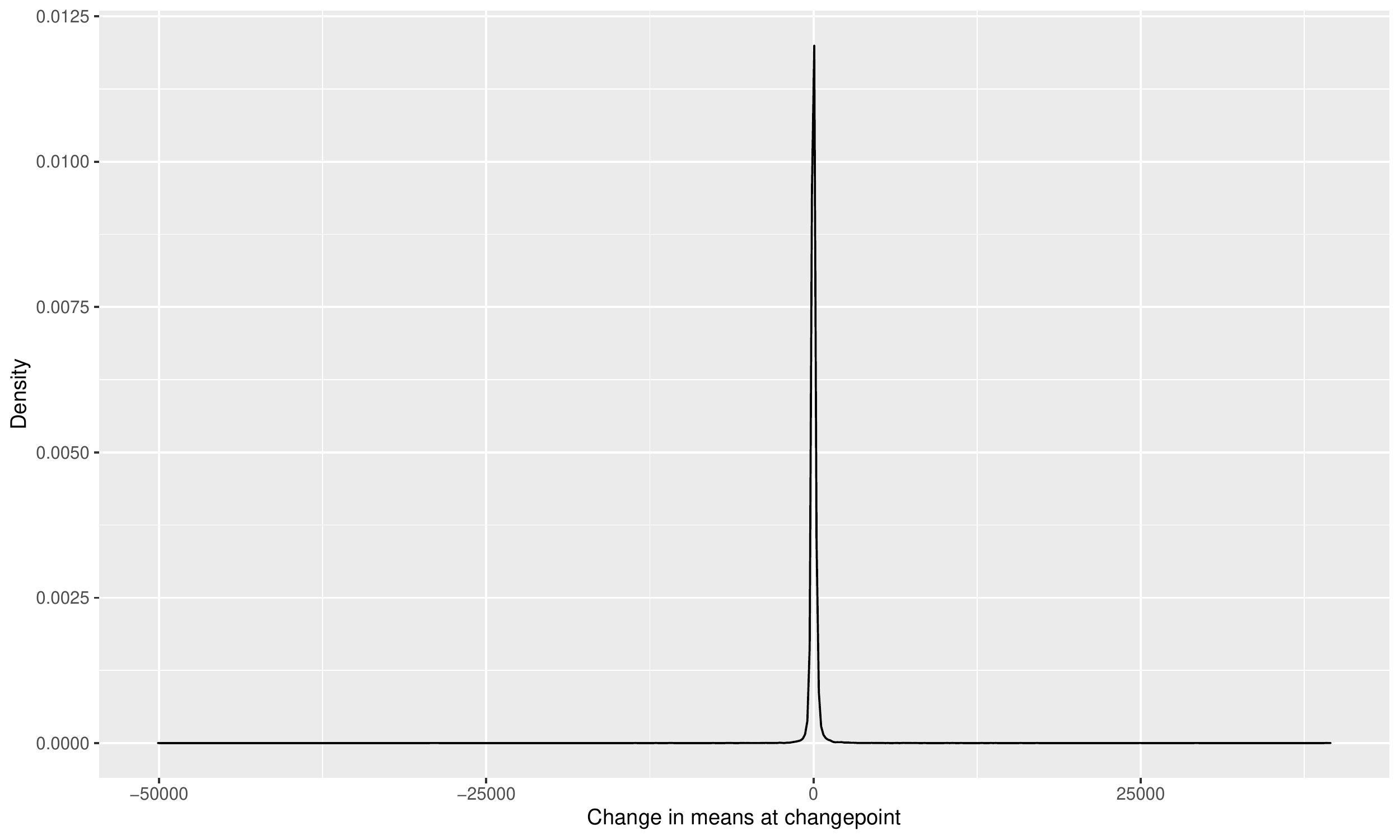}
    \caption{Size of Changes in Commit Time Series}
    \label{fig:commitsizes}
\end{figure}

We found a total of 31,416 changepoints in commit time series, of which 15,342 (49\%) were increases in commit activity and 16,047 (51\%) were reductions in activity. We computed the magnitude of a changepoint as the difference in means in the number of monthly commits before and after the changepoint. The size of most changes were relatively small, with an interquartile range (IQR) of -75 to 87 commits per month, but there was a substantial tail in both directions as can be seen in Figure~\ref{fig:commitsizes}.

We found 28,671 changepoints in author time series, of which 12,114 (42\%) were reductions and 16,557 (58\%) were increases in activity. Changes in the number of contributing authors per month were relatively small, with an IQR between -3.4 to 5.8 authors per month, but there was again a substantial tail in both directions. The graph for author time series is identical in appearance to Figure~\ref{fig:commitsizes} though the scale differs.

\section{Conclusion}

We found that open source evolution is rarely smooth and typically includes changepoints. The vast majority of projects had between one and six changepoints in both project activity time series, though some outliers had up to 16 changepoints. Increases and decreases in activity occurred with roughly equal frequency. While most changepoints are relatively small (a few authors per month, a few dozen commits per month), there is a long tail of much larger changes. The data and code used in this project can be found in the project's git repository at \url{https://github.com/woc-hack/inflection-points}.

In the future, we plan to study patterns of changepoints in an attempt to identify common lifecycle models and common responses to external events, such as security incidents~\cite{openssl2020}. We plan to focus first on changepoints with large differences in the mean, as these are more likely to indicate major changes in project direction. We also plan to examine changepoints in software characteristics beyond project activity, such as code size and complexity. 

\bibliographystyle{IEEEtran}
\bibliography{changepoints-msr2021}

\end{document}